\documentclass[aps,twocolumn,pra,superscriptaddress,showpacs,tightenlines]{revtex4}
\usepackage{amssymb}
\usepackage{amsmath}
\usepackage{graphicx}
\usepackage{epsfig}
\usepackage{txfonts}
\usepackage{subfigure}
\usepackage{amsfonts}
\usepackage{CJK}

\begin{document}

\title{Controllable coupling and quantum correlation dynamics of two double quantum dots \\coupled via a transmission line resonator}
\author{\rm Qin-Qin Wu }
\affiliation{Laboratory of Low-Dimensional Quantum Structures and
Quantum Control of Ministry of Education, and Department of
Physics, Hunan Normal University, Changsha 410081,
China}\affiliation{Department of Physics and electronics, Hunan
Institute of Science and Technology, Yueyang 414000, China}
\author{ Qing-Shou Tan}
\affiliation{Laboratory of Low-Dimensional Quantum Structures and
Quantum Control of Ministry of Education, and Department of
Physics, Hunan Normal University, Changsha 410081, China}
\author{Le-Man Kuang\footnote{Author to whom any correspondence should be
addressed. Email: lmkuang@hunnu.edu.cn}} \affiliation{Laboratory
of Low-Dimensional Quantum Structures and Quantum Control of
Ministry of Education, and Department of Physics, Hunan Normal
University, Changsha 410081, China}
\date{\today}

\begin{abstract}
We propose a theoretical scheme to generate a controllable and
switchable coupling between two double-quantum-dot (DQD) spin qubits
by using a transmission line resonator (TLR) as a bus system. We
study dynamical behaviors of quantum correlations described by
entanglement correlation (EC) and discord correlation (DC) between
two DQD spin qubits when the two spin qubits and the TLR are
initially prepared in $X$-type quantum states and a coherent state,
respectively. We demonstrate that in the EC death regions there
exist DC stationary states in which the stable DC amplification or
degradation can be generated during the dynamical evolution. It is
shown that these DC stationary states can be controlled by
initial-state parameters, the coupling, and detuning between qubits
and the TLR. We reveal the full synchronization and
anti-synchronization phenomena in the EC and DC time evolution, and
show that the EC and DC synchronization and anti-synchronization
depends on the initial-state parameters of the two DQD spin qubits.
These results shed new light on dynamics of quantum correlations.
\end{abstract}
\maketitle

\section{Introduction}

It is well known that quantum entanglement \cite{ami,hor} and
quantum discord \cite{31} are two different types of quantum
correlations. In recent years, it has been widely recognized that
both of the two quantum correlations are essential quantum
resources which can be used to realize quantum information
processing, and the fact that quantum discord is a more general
concept to measure quantum correlations than quantum entanglement
since there is a nonzero quantum discord in some separable mixed
states \cite{25,26,27,28,29,30,32,33}. However, the question of
the relation between entanglement correlation (EC) and discord
correlation (DC) is still an open problem in the field of quantum
correlations.

Interactions of quantum systems are at the core of quantum
information processing. In particular, quantum computing requires
that inter-qubit interactions are controllable, and can be
selectively switched on and off \cite{mak,liu}. Many physical
systems have been explored for the realization of practical quantum
information processors, such as cavity quantum electrodynamics (QED)
system \cite{1}, optical system \cite{2}, and solid-state system
\cite{3,4,ble,chi,sch,kub,wu}. Solid-state devices are the promising
candidates for the implementation of quantum computation due to the
possibility of fabricating large integrated networks. Among
different kinds of solid systems, double quantum dot (DQD) system
\cite{5,6,7,8,9,10,11,12,13} and a transmission line resonator (TLR)
\cite{14,15,16,17,18,19,20} are particularly attractive because of
the relative long spin coherence time and high controllability of
DQD system and quantum bus function of the TLR. Several proposals
have been proposed to realize controllable couplings and local
operations of quantum dot (QD) qubits via a TLR \cite{21,24,22,23}.
In Ref. \cite{21}, the electron spins in nanowire QDs couple to the
electric component of the resonator electromagnetic field and enable
quantum information processing in an all-electrical fashion. In Ref.
\cite{24}, all-electrical coupling between QD spin qubits and a TLR
are also used to produce effective interactions between spin qubits.
In Ref. \cite{22}, two QD spin qubits were embedded in a
superconducting microstrip cavity, virtual photons in a common
cavity mode could mediate coherent interactions between two distant
qubits.  In this paper, we want to propose a new scheme to implement
the controllable coupling between double-quantum-dot (DQD) spin
qubits in terms of the magnetic coupling between DQDs  and the TLR
by using the TLR as the quantum bus system. In our scheme, each DQD
spin qubit couples to the TLR through the magnetic filed generated
by the current of the TLR. We will study the relation between EC and
DC by investigating the dynamic behaviors of  EC and DC between the
two DQD spin qubits in the combining system consisting of two DQDs
and a TLR. We will show the appearance of the DC stationary states
in the EC death regimes and demonstrate the full synchronization and
anti-synchronization of DC and EC in the time evolution. In
particular, it shall be indicated that the DC between the two DQD
spin qubits can reach and keep its maximum fixed even if the EC
disappears completely in the time evolution for certain initially
prepared $X$-type states.

This paper is organized as follows. In Sec.~\ref{Sec:2}, we
present our coupling scheme of two DQD spin qubits by using the
TLR as the quantum bus system, and show that the inter-spin-qubit
interaction in our coupling scheme is controllable and switchable.
In Sec.~\ref{Sec:3}, we investigate dynamical behaviors of EC and
DC  between two DQD qubits in the controllable coupling scheme,
and discuss the relation between EC and DC for $X$-type initial
states. The stationary amplification and degradation of the DC and
the time synchronization and anti-synchronization of the EC and DC
are revealed in the time evolution of the combined hybrid system.
Finally, we conclude this work in Sec.~\ref{Sec:5}.

\section{\label{Sec:2} Controllable Coupling scheme of two DQD spin qubits}

In this section, we propose a scheme to generate an effective
controllable interaction between two DQD spin qubits by using a TLR
as the data bus.  The combined system under our consideration is
indicated in Fig. 1. It consists of two DQDs charged with two excess
electrons and a TLR. The length of the TLR is $L$. The two DQDs, 1
and 2, are placed at the positions $L/4$ and $3L/4$ of the TLR,
respectively. These are the antinode of the quantized current of the
TLR. Two electron spins in each DQD are localized in adjacent QDs,
coupled via tunnelling. The distance between the lower dot and the
TLR and the distance between the two dots are  $r$.

In terms of the annihilation and creation operators $a$ and
$a^{\dag}$, we can write the Hamiltonian for the TLR as
\begin{eqnarray}
\label{e1}
 \hat{H}_{r}=\hbar\omega_{r}\hat{a}^{\dag}\hat{a},
\end{eqnarray}
where $\omega_{r}$ is the frequency of the TLR. The Hamiltonian of a
DQD is most conveniently written in the two-electron singlet-triplet
basis
$\{|S_{11}\rangle,|T_{11}^{0}\rangle,|T_{11}^{+}\rangle,|T_{11}^{-}\rangle,|S_{02}\rangle\}$
with the quantization axis in the $z$-direction as
\cite{23,34,35,36}
\begin{eqnarray} \label{e2}
\hat{H}_{q}&=&E_{S}|S_{11}\rangle\langle
S_{11}|+(\Delta_{0}+E_{S})|S_{02}\rangle\langle
S_{02}|\\&&+g_{B}\mu_{B}B_{e}(|T_{11}^{+}\rangle\langle
T_{11}^{+}|-|T_{11}^{-}\rangle\langle T_{11}^{-}|)\nonumber\\
&&+E_{T}|T_{11}^{0}\rangle\langle
T_{11}^{0}|+t(|S_{11}\rangle\langle S_{02}|+|S_{02}\rangle\langle
S_{11}|),\nonumber
\end{eqnarray}
where the two-electron singlet-triplet basis are given by
\begin{eqnarray}
\label{e3} |T_{11}^{+}\rangle&=&|\uparrow\uparrow\rangle, \;\;\;
|T_{11}^{0}\rangle
=\frac{1}{\sqrt{2}}(|\uparrow\downarrow\rangle+|\downarrow\uparrow\rangle),
\;\;\;|T_{11}^{-}\rangle=|\downarrow\downarrow\rangle,  \nonumber \\
|S_{11}\rangle&=&\frac{1}{\sqrt{2}}(|\uparrow\downarrow\rangle-|\downarrow\uparrow\rangle),
\end{eqnarray}
and the auxiliary singlet state with two electrons in one quantum dot,
$|S_{02}\rangle$, is coupled via tunneling $t$ to the separated singlet, $|S_{11}\rangle$,
where the subscript $m$, $n$ denotes the dot occupancy. $E_{S}$ and $E_{T}$ are the energy of the $|S_{11}\rangle$ and $|T_{11}^{0}\rangle$ states, respectively.
$\Delta_{0}$
is the energy difference between the $|S_{11}\rangle$ and
$|S_{02}\rangle$ states set by the electric field. $B_{e}$ is the external magnetic field, $\mu_{B}$ is
the Bohr magneton and $g_{B}$ the electron spin $g$-factor.

\begin{figure} [htp]
\center
\includegraphics[bb=44 632 238 772 width=3.0in] {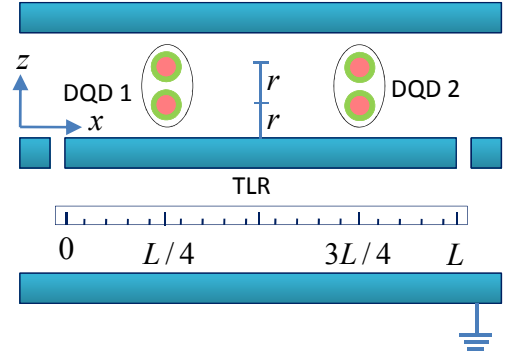}
\caption{The proposed setup with two DQDs, biased with external
potential $\Delta_{0}^{j},\;j=1,2$, magnetically coupled to a TLR of
length $L$. The two DQDs 1 and 2 are placed at the positions $L/4$
and $3L/4$ of the TLR, respectively. These are the antinode of the
quantized current of the TLR. In each DQD, both of the distance
between the lower dot and the TLR and the distance between the two
dots are $r$.}
\end{figure}

If the external magnetic field in the upper dot $B_{e}^{u}$
 is different from that in the lower dot $B_{e}^{l}$, and if we consider the effect of the
hyperfine interaction with the nuclear spins which can be studied by
adding a static frozen effective nuclear field $B_{N}^{u}$
($B_{N}^{l}$) at the upper (lower) dot to the total magnetic field
\cite{34,35,36}, the interaction between the spin and  the total
magnetic field in each DQD is given by
$\hat{H}_{I}=-g_{B}\mu_{B}(\hat{B}_{u}\cdot\hat{S}_{u}+\hat{B}_{l}\cdot\hat{S}_{l})/\hbar$
 which can be rewritten as
\begin{eqnarray}
\label{e4}
\hat{H}_{I}&=&-\frac{g_{B}\mu_{B}}{\hbar}(\hat{B}_{u}+\hat{B}_{l})\cdot(\hat{S}_{u}+\hat{S}_{l})/2,\nonumber\\
&&-\frac{g_{B}\mu_{B}}{\hbar}(\hat{B}_{u}-\hat{B}_{l})\cdot(\hat{S}_{u}-\hat{S}_{l})/2,
\end{eqnarray}
where the total magnetic fields which an electron in each DQD
experiences are $\hat{B}_{u}=\hat{B}_{N}^{u}+\hat{B}_{e}^{u}$
 and $\hat{B}_{l}=\hat{B}_{N}^{l}+\hat{B}_{e}^{l}$, the sum operators and the
 difference operators of the two electron spins can be expressed in terms of the two-electron singlet-triplet basis as
\begin{eqnarray}
\label{e6}
\hat{S}_{u}^{x}+\hat{S}_{l}^{x}&=&\frac{\hbar}{\sqrt{2}}(|T_{11}^{0}\rangle\langle
T_{11}^{-}|+|T_{11}^{0}\rangle\langle T_{11}^{+}|+ \texttt{H.c.}),\nonumber\\
\hat{S}_{u}^{y}+\hat{S}_{l}^{y}&=&\frac{\hbar}{\sqrt{2}}(i|T_{11}^{0}\rangle\langle
T_{11}^{-}|-i|T_{11}^{0}\rangle\langle T_{11}^{+}|+ \texttt{H.c.}),\nonumber\\
\hat{S}_{u}^{z}+\hat{S}_{l}^{z}&=&\hbar(|T_{11}^{+}\rangle\langle
T_{11}^{+}|-|T_{11}^{-}\rangle\langle T_{11}^{-}|),\nonumber\\
\hat{S}_{u}^{x}-\hat{S}_{l}^{x}&=&\frac{\hbar}{\sqrt{2}}(|S_{11}\rangle\langle
T_{11}^{-}|-|S_{11}\rangle\langle T_{11}^{+}|+ \texttt{H.c.}),\nonumber\\
\hat{S}_{u}^{y}-\hat{S}_{l}^{y}&=&\frac{\hbar}{\sqrt{2}}(i|S_{11}\rangle\langle
T_{11}^{-}|+i|S_{11}\rangle\langle T_{11}^{+}|+ \texttt{H.c.}),\nonumber\\
\hat{S}_{u}^{z}-\hat{S}_{l}^{z}&=&\hbar(|S_{11}\rangle\langle
T_{11}^{0}|+|T_{11}^{0}\rangle\langle S_{11}|).
\end{eqnarray}
Eqs. (\ref{e4}) and (\ref{e6}) indicate that the homogeneous part of
the magnetic field $\hat{B}_{u}+\hat{B}_{l}$ simply adds vectorially
to the external field $B_{e}$, changing slightly the Zeeman
splitting and preferred spin orientation of the triplet states. The
inhomogeneous part $\Delta \hat{B}=\hat{B}_{u}-\hat{B}_{l}$, on the
other hand, couples the triplet states
$\{|T_{11}^{0}\rangle,|T_{11}^{+}\rangle,|T_{11}^{-}\rangle\}$ to
the singlet state $|S_{11}\rangle$.

The degree of mixing between two states will depend strongly on the
energy difference between them. In the case of $g_{B}\mu_{B}B_{e}\gg
t$ and  $g_{B}\mu_{B}\sqrt{\langle \Delta B^{2}\rangle}$, the
spin-aligned states $|T_{11}^{+}\rangle$ and $|T_{11}^{-}\rangle$
are split off due to Zeeman energy $g_{B}\mu_{B}B_{e}$, then the
perturbation of these states will be small, while the
spin-anti-aligned state $|T_{11}^{0}\rangle$ remains mixed with the
state  $|S_{11}\rangle$. Then we can write the Hamiltonian of this
combined system as
\begin{eqnarray}
\label{e7}
\hat{H}&=&\hbar\omega_{r}\hat{a}^{\dag}\hat{a}+\hbar\sum_{j=1}^{2}\frac{\omega_{j}}{2}\hat{\sigma}_{z}^{j}
+(\Delta_{0}^{j}+E_{S}^{j})|S_{02}\rangle_{j}\langle
S_{02}|\nonumber\\&+&\frac{1}{2}g_{B}\mu_{B}
(\Delta \hat{B}_{j}^{z})\hat{\sigma}_{x}^{j}+t_{j}(|S_{11}\rangle_{j}\langle
S_{02}|+|S_{02}\rangle_{j}\langle S_{11}|),
\end{eqnarray}
where $\omega_{j}=(E_{S}^{j}-E_{T}^{j})/\hbar$, $\Delta
\hat{B}_{j}^{z}$ is the $z$-component of $\Delta \hat{B}_{j}$, and
we have introduced Puali spin operators
\begin{eqnarray}
\label{e8} \sigma_{z}^{j}&=&|S_{11}\rangle_{j}\langle
S_{11}|-|T_{11}^{0}\rangle_{j}\langle
T_{11}^{0}|,\nonumber\\
\sigma_{x}^{j}&=&|S_{11}\rangle_{j}\langle
T_{11}^{0}|+|T_{11}^{0}\rangle_{j}\langle S_{11}|.
\end{eqnarray}

In Hamiltonian (\ref{e7})  we have used only
$|T_{11}^{0}\rangle_{j}$ and $|S_{11}\rangle_{j}$ as a spin-qubit
degree of freedom. Qubit-TLR Interaction is induced naturally by the
magnetic field gradient $\Delta \hat{B}_{j}^{z}$ which includes the
contributions from the nuclear magnetic field and the TLR itself
\begin{eqnarray} \label{e9}
\Delta \hat{B}_{j}^{z}=\Delta \hat{B}_{N_{j}}^{z}+\frac{\mu_{0}\hat{I}_{j}}{4\pi r},
\end{eqnarray}
where $\mu_{0}$ is the vacuum permeability, and $\Delta
\hat{B}_{N_{j}}^{z}$ is the gradient of the longitudinal component
of the nuclear magnetic fields between the two dots of the $j$th
DQD. The current $\hat{I}_{j}$ at the position of the $j$th DQD due
to the resonator of length $L$  is quantized as
\begin{eqnarray} \label{e10}
\hat{I}_{j}=(-1)^{j-1}\sqrt{\frac{
\hbar\omega_{r}}{Ll}}(\hat{a}+\hat{a}^{\dag}),
\end{eqnarray}
where $l$ is the inductance per unit length of the TLR.

Substituting Eqs. (\ref{e9}) and (\ref{e10})  into Eq. (\ref{e7}) we
arrive at the following Hamiltonian
\begin{eqnarray} \label{e11}
\hat{H}&=&\hbar\omega_{r}\hat{a}^{\dag}\hat{a}+\hbar\sum_{j=1}^{2}\frac{
\omega_{j}}{2}\hat{\sigma}_{z}^{j}
+(\Delta_{0}^{j}+E_{S}^{j})|S_{02}\rangle_{j}\langle
S_{02}|\nonumber\\
&&+\frac{g_{B}\mu_{B}}{2}\left[\Delta \hat{B}_{N_{j}}^{z}+(-1)^{j-1}\frac{\mu_{0}}{4\pi r}\sqrt{\frac{\hbar\omega_{r}}{Ll}}(\hat{a}+\hat{a}^{\dag})\right]\hat{\sigma}_{x}^{j}\nonumber\\
&&+t_{j}(|S_{11}\rangle_{j}\langle
S_{02}|+|S_{02}\rangle_{j}\langle S_{11}|).
\end{eqnarray}

In the interaction picture with respect to the term
$(\Delta_{0}^{j}+E_{S}^{j})|S_{02}\rangle_{j}\langle S_{02}|$, we
obtain the interaction Hamiltonian after discarding rapidly
oscillating terms
\begin{eqnarray} \label{e12}
\hat{H}_{1}&=&\hbar\omega_{r}\hat{a}^{\dag}\hat{a}+\hbar\sum_{j=1}^{2}\frac{
\omega_{j}}{2}\hat{\sigma}_{z}^{j}\\
&&+\frac{g_{B}\mu_{B}}{2}\left[\Delta \hat{B}_{N_{j}}^{z}+(-1)^{j-1}\frac{\mu_{0}}{4\pi r}\sqrt{\frac{\hbar\omega_{r}}{Ll}}(\hat{a}+\hat{a}^{\dag})\right]\hat{\sigma}_{x}^{j}.\nonumber
\end{eqnarray}

When  $\omega_{j}\gg g_{B}\mu_{B}\Delta B_{N_{j}}^{z}/(2\hbar)$, and
$\omega_{j}+\omega_{r}\gg\omega_{j}-\omega_{r}$,
$\frac{g_{B}\mu_{B}\mu_{0}}{8\hbar\pi
r}\sqrt{\frac{\hbar\omega_{r}}{Ll}}$, the rotating-wave
approximation can be applied to get an effective Hamiltonian of the
combined system
\begin{eqnarray} \label{e13}
\hat{H}_{2}=\hbar\omega_{r}\hat{a}^{\dag}\hat{a}+\hbar\sum_{j=1}^{2}\frac{
\omega_{j}}{2}\hat{\sigma}_{z}^{j}+ (-1)^{j-1} \hbar
g(a\hat{\sigma}_{+}^{j}+\hat{\sigma}_{-}^{j}a^{\dag}),
\end{eqnarray}
where the effective coupling constant
$g=\frac{g_{B}\mu_{B}\mu_{0}}{8\hbar\pi
r}\sqrt{\frac{\hbar\omega_{r}}{Ll}}$. Eq. (\ref{e13}) is the
Hamiltonian of the usual Jaynes-Cummings model of two atoms with
$\hat{\sigma}_{+}^{j}=|S_{11}\rangle_{j}\langle T_{11}^{0}|$ and
$\hat{\sigma}_{-}^{j}=|T_{11}^{0}\rangle_{j}\langle S_{11}|$.

If  both DQDs are strongly detuned from the TLR, i.e.,
$|\delta_{j}|=|\omega_{j}-\omega_{r}|\gg |g|$, we can adiabatically
eliminate the TLR mode using the following transformation
\begin{eqnarray}
\label{e14}
\hat{U}=\exp\left[\frac{g}{\delta_{1}}(\hat{a}^{\dag}\hat{\sigma}_{1}^{-}-\hat{a}\hat{\sigma}_{1}^{+})-\frac{g}{\delta_{2}}(\hat{a}^{\dag}\hat{\sigma}_{2}^{-}-\hat{a}\hat{\sigma}_{2}^{+})\right].
\end{eqnarray}
To second order in the small parameters $g/\delta_{j}$, the
effective Hamiltonian becomes
\begin{eqnarray} \label{e15}
\hat{H}_{\texttt{eff}}&=&\hbar\omega_{r}\hat{a}^{\dag}\hat{a}+\frac{\hbar}{2}\sum_{j=1}^{2}\left[\omega_{j}+2\frac{g^{2}}{\delta_{j}}\left(\hat{a}^{\dag}\hat{a}+\frac{1}{2}\right)\right]\hat{\sigma}_{z}^{j}\nonumber\\
& &-\frac{\hbar
g^{2}(\delta_{1}+\delta_{2})}{2\delta_{1}\delta_{2}}(\hat{\sigma}_{+}^{1}\hat{\sigma}_{-}^{2}+\hat{\sigma}_{-}^{1}\hat{\sigma}_{+}^{2}),
\end{eqnarray}
where the last term  describes swap of the DQD states through
virtual interaction with the TLR.

To easily switch on and off the inter-qubit coupling is one of the
most important open problems in quantum computing hardware. Here we
propose a way to overcome the severe problem.  From Eqs. (\ref{e9}),
(\ref{e13}) and (\ref{e15}) we can see that the inter-spin-qubit
coupling between the two DQDs is mediated by the TLR, and the
interaction between each DQD and the TLR originated from the
magnetic field gradient $\mu_{0}I_{j}/4\pi r$ between the two dots
of each DQD, which is produced by the TLR. So if additionally we add
an asymmetric magnetic field $B_{z}$ in the $z$-direction, then we
have
\begin{eqnarray} \label{e16}
\Delta B_{j}^{z}=\Delta B_{N_{j}}^{z}+\frac{\mu_{0}I_{j}}{4\pi r}+d B_{z}^{j},
\end{eqnarray}
where $d B_{z}^{j}$ is the difference of the asymmetric magnetic
field $B_{z}$ between the two QDs of the $j$th DQD. Then by tuning
the magnitude of $d B_{z}^{j}$, we can get $\Delta B_{j}^{z}=0$.
This implies that the interaction between the $j$th DQD and the TLR
is turned off. Therefore, we can conclude that the effective
coupling between the two DQDs in the present spin-qubit coupling
scheme is controllable and switchable.

\section{\label{Sec:3} Dynamics of quantum correlations between two DQD spin qubits}

In this section, we investigate dynamics of quantum correlations
between two DQD spin qubits. We will study dynamic evolution of
quantum entanglement and quantum discord when the two DQD spin
qubits are initially in three-parameter two-qubit $X$-type quantum
states  \cite{yu} which play an important role in a number of
physical systems.

Firstly, we solve the Hamiltonian of the DQD-TLR system under our
consideration. In the interaction picture with respect to the first
term $\hbar\omega_{r}\hat{a}^{\dag}\hat{a}$ of the Hamiltonian
$\hat{H}_{\texttt{eff}}$ (\ref{e15}), we get
\begin{eqnarray} \label{e17}
\hat{\mathcal{H}}=\hbar\sum_{j=1}^{2}\Omega_{j}\hat{\sigma}_{z}^{j}-\hbar\chi(\hat{\sigma}_{+}^{1}\hat{\sigma}_{-}^{2}+\hat{\sigma}_{-}^{1}\hat{\sigma}_{+}^{2}).
\end{eqnarray}
where the inter-qubit and qubit-TLR couplings are given by
\begin{eqnarray} \label{e18}
\Omega_{j}&=&\frac{1}{2}\left[\omega_{j}+2\frac{g^{2}}{\delta_{j}}\left(\hat{N}+\frac{1}{2}\right)\right],\;\;\;
\chi=\frac{g^{2}(\delta_{1}+\delta_{2})}{2\delta_{1}\delta_{2}},
\end{eqnarray}
with $\hat{N}=a^{\dag}a$ being the number operator of the TLR. We
can write $\hat{\mathcal{H}}$ in the basis $\{\vert
S\rangle_{1}\vert S\rangle_{2},\vert S\rangle_{1}\vert
T\rangle_{2},\vert T\rangle_{1}\vert S\rangle_{2},\vert
T\rangle_{1}\vert T\rangle_{2}\}$ as
\begin{eqnarray}
\label{e19} \hat{\mathcal{H}}=\hbar\left(
\begin{array}{cccc}
\Omega_{1}+\Omega_{2} & 0 & 0 & 0 \\
0 & \Omega_{1}-\Omega_{2} & -\chi & 0 \\
0 & -\chi & -\Omega_{1}+\Omega_{2} & 0 \\
0 & 0 & 0 & -\Omega_{1}-\Omega_{2}
\end{array}
\right).
\end{eqnarray}
Here and in the after we use $|S\rangle_{j}$ and $|T\rangle_{j}$ to
replace $\vert S_{11}\rangle_{j}$ and $\vert T_{11}^{0}\rangle_{j}$
in Sec.~\ref{Sec:2}, respectively. The four eigenstates of
the Hamiltonian $\hat{\mathcal{H}}$ can be obtained as
\begin{eqnarray}
 \label{e20}
\vert \Psi _{1}\rangle &=&\vert S\rangle_{1}\vert S\rangle_{2}, \nonumber \\
\vert \Psi _{2}\rangle &=&\cos(\theta/2)\vert S\rangle_{1}\vert
T\rangle_{2} +\sin(\theta/2)\vert T\rangle_{1}\vert S\rangle_{2}, \nonumber \\
\vert \Psi_{3}\rangle &=&-\sin(\theta/2)\vert S\rangle_{1}\vert
T\rangle_{2} +\cos (\theta/2)\vert T\rangle_{1}\vert S\rangle_{2}, \nonumber\\
\vert \Psi_{4}\rangle &=&\vert T\rangle_{1}\vert T\rangle_{2},
\end{eqnarray}
with the corresponding eigenvalues
\begin{eqnarray}
\label{e21}
E_{1}&=&-E_{4}=\hbar(\Omega_{1}+\Omega_{2}), \nonumber \\
E_{2}&=&-E_{3}=\hbar\sqrt{(\Omega_{1}-\Omega_{2})^{2} +\chi^{2}}.
\end{eqnarray}
The mixing angle in Eq. (\ref{e20}) is defined by
\begin{eqnarray}
\label{e22}
\sin\theta=-\hbar\chi/E_{2},\;\;\;\;\;\cos\theta=\hbar(\Omega_{1}-\Omega_{2})/E_{2}.
\end{eqnarray}

From Eqs. (\ref{e20}) and (\ref{e21}) we get the time evolution
operator of $\hat{\mathcal{H}}$ as $\hat{U} =\sum_{n=1}^{4}\exp
(-iE_{n}t/\hbar)\vert \Psi _{n}\rangle \langle \Psi _{n}\vert $
which can be expressed as
\begin{eqnarray}
 \label{e23}
\hat{U} &=&\left(
\begin{array}{cccc}
e^{-iE_{1}t/\hbar} & 0 & 0 & 0 \\
0 & e^{-iE_{2}t/\hbar}-\eta
& \kappa & 0 \\
0 & \kappa &
e^{-iE_{3}t/\hbar}+\eta & 0 \\
0 & 0 & 0 & e^{-iE_{4}t/\hbar}
\end{array}
\right)
\end{eqnarray}
with the following parameters
\begin{eqnarray}
 \label{e24}
\kappa&=&(e^{-iE_{2}t/\hbar}-e^{-iE_{3}t/\hbar})\sin(\theta/2)\cos(\theta/2),\nonumber\\
\eta&=&(e^{-iE_{2}t/\hbar}-e^{-iE_{3}t/\hbar})\sin^{2}(\theta/2).
\end{eqnarray}

We assume the two DQDs are initially prepared in a class of state
with maximally mixed marginals ($\hat{\rho}
_{A(B)}=\hat{I}_{A(B)}/2$) described by the three-parameter $X$-type
density matrix
$\hat{\rho}(0)=\left(\hat{I}_{AB}+\sum_{i=1}^{3}c_{i}\hat{\sigma}_{A}^{i}\otimes
\hat{\sigma} _{B}^{i}\right)/4$ which can be expressed as
\begin{eqnarray}
\label{25} \hat{\rho}(0) &=&\frac{1}{4}\left(
\begin{array}{cccc}
1+c_{3} & 0 & 0 & c_1-c_2 \\
0 & 1-c_{3} & c_1+c_2 & 0 \\
0 & c_1+c_2 & 1-c_{3} & 0 \\
c_1-c_2 & 0 & 0 & 1+c_{3}
\end{array}
\right),
\end{eqnarray}
where $c_{i}$ ($0\leq \left\vert c_{i}\right\vert \leq 1$) are real
numbers satisfying the unit trace and positivity conditions of the
density operator $\hat{\rho}$. The density operator $\hat{\rho}$
includes the Werner states and the Bell states as two special cases.
Here and in the after we use the label $A$($B$) to denote the first
(second) DQD.

If we initially prepare the TLR in the coherent state
\begin{eqnarray}
 \label{e26}
 \vert\psi_{r}(0)\rangle
 =\vert\alpha\rangle_{r}=e^{-\frac{1}{2}|\alpha|^2}\sum^{\infty}_{n=0}\frac{\alpha^n}{\sqrt{n!}}|n\rangle.
\end{eqnarray}

Then, the reduced density matrix of the two DQDs at time $t$ can be
obtained as
\begin{eqnarray}
 \label{e27}
\hat{\rho}(t)&=&\texttt{Tr}_{r}(\hat{U}^{\dagger
}\hat{\rho}(0)\otimes \vert \psi_{r} (0)\rangle\langle
\psi_{r}(0)\vert \hat{U}).
\end{eqnarray}
For simplicity but without loss generality, here we consider the
case of two identical DQDs, i.e., we take
$\omega_{A}=\omega_{B}=\omega$ and $\delta_{A}=\delta_{B}=\delta$.
Then from Eqs. (\ref{e18}) and (\ref{e22}) we have $\cos\theta=0$.
In this case, the explicit form of density operator at time $t$ is
\begin{eqnarray}
 \label{e28}
\hat{\rho}(t) &=&\frac{1}{4}\left(\begin{array}{llll}
1+c_{3} & 0 & 0 & c_0\\
0 & 1-c_{3} & c_{1}+c_{2} & 0 \\
0 & c_{1}+c_{2} & 1-c_{3} & 0 \\
c^*_0& 0 & 0 & 1+c_{3}
\end{array}
\right),
\end{eqnarray}
where we have introduced the following parameter
\begin{eqnarray}
\label{e29}c_0&=&(c_{1}-c_{2})e^{2i(\omega+g^{2}/\delta)t}\exp{\left[-|\alpha|^{2}\left(1-e^{4i
g^{2}t/\delta}\right)\right]}.
\end{eqnarray}

In what follows, we study dynamic properties of  quantum
correlations between the two DQD spin qubits in terms of the
expression of the density operator of the two DQD spin qubits given
in Eq. (\ref{e28}). We investigate in detail EC and DC described by
concurrence \cite{woo} and quantum discord, respectively. When the
density matrix of the two qubit system has an $X$-type structure,
the concurrence has a simple analytic expression \cite{yu}
\begin{eqnarray}
\label{e30} C(t)_{AB}=2\max\{0,\Lambda_{1}(t),\Lambda_{2}(t)\},
\end{eqnarray}
where $\Lambda _{1}(t)=\vert \rho _{14}(t)\vert -\sqrt{\rho
_{22}(t)\rho _{33}(t)}$ and $\Lambda _{2}(t)=\vert \rho
_{23}(t)\vert -\sqrt{\rho _{11}(t)\rho _{44}(t)}$ with
$\rho_{ij}\;(i,j=1,2,3,4)$ being the matrix elements of the density
operator $\hat{\rho}(t)$ in Eq. (27). For the density operator given
by Eq. (27) we have
\begin{eqnarray}
\label{e31} \Lambda _{1}(t)&=&\frac{1}{4}(|c_0|-|1-c_{3}|),\nonumber\\
 \Lambda _{2}(t)&=&\frac{1}{4}(|c_{1}+c_{2}|-|1+c_{3}|).
\end{eqnarray}

Now we turn to investigate dynamic evolution of quantum discord
correlation between two DQD qubits under our consideration. Quantum
discord \cite{31} is defined as the difference between the total
correlation and the classical correlation with the following
expression
\begin{equation}
\label{32} \mathcal {D}\left( \hat{\rho}\right) =\mathcal{I}\left(
\hat{\rho}_{A}:\hat{\rho}_{B}\right)
-\mathcal{C}\left(\hat{\rho}\right).
\end{equation}
Here the total correlation in a bipartite quantum state $\hat{\rho}$
is measured by quantum mutual information given by
\begin{eqnarray}
\label{33} \mathcal{I}\left( \hat{\rho} _{A}:\hat{\rho} _{B}\right)
=S\left( \hat{\rho} _{A}\right) +S\left(\hat{\rho} _{B}\right)
-S\left( \hat{\rho}\right),
\end{eqnarray}
where $S\left( \hat{\rho} \right) =-\textrm{Tr}(\hat{\rho} \log
\hat{\rho})$ is the von Neumann entropy,
$\hat{\rho}_{A}=\textrm{Tr}_{B}(\hat{\rho})$ and
$\hat{\rho}_{B}=\textrm{Tr}_{A}(\hat{\rho})$ are  the reduced
density operators for subsystems $A$ and $B$, respectively. And the
classical correlation between the two subsystems $A$ and $B$ can be
defined as
\begin{eqnarray}
\label{34} \mathcal{C}(\hat{\rho})&=&\max_{\{\hat{P}_{k}\}}\left[S(\hat{\rho}
_{A})-\sum_{k}p_{k}S\left(\hat{\rho}_{A}^{(k)}\right)\right]\nonumber\\
&=&S(\hat{\rho} _{A})-\min_{\{\hat{P}_{k}\}}\left[
\sum_{k}p_{k}S\left(\hat{\rho}_{A}^{(k)}\right)\right]. \label{clacor}
\end{eqnarray}
Here $\{\hat{P}_{k}\}$ is a set of projects performed locally on the
subsystem $B$, and $\hat{\rho}
_{A}^{(k)}=\frac{1}{p_{k}}\textrm{Tr}_{B}\left[\left(
\hat{I}_{A}\otimes
\hat{P}_{k}\right)\hat{\rho}\left(\hat{I}_{A}\otimes
\hat{P}_{k}\right)\right]$ is the state of the subsystem $A$
conditioned on the measurement outcome labeled by $k$ ,where
$p_{k}=\textrm{Tr}_{AB}\left[\left(\hat{I}_{A}\otimes
\hat{P}_{k}\right)\hat{\rho}\left(\hat{I}_{A}\otimes
\hat{P}_{k}\right)\right]$ denotes the probability relating to the
outcome $k$, and $\hat{I}_{A}$ denotes the identity operator for the
subsystem $A$.

In order to obtain the quantum discord for the two qubit-system
\cite{yuan}, we first evaluate the mutual information of state
$\hat{\rho}(t)$ given in Eq. (\ref{e28}). The four eigenvalues of
$\hat{\rho}(t)$ are
\begin{eqnarray}\label{e35}
\lambda _{1,2}
&=&\frac{1}{4}(1-c_{3}\pm\vert c_{1}+c_{2}\vert ),\nonumber\\
\lambda _{3,4}&=&\frac{1}{4}(1+c_{3}\pm\vert c_0\vert).
\end{eqnarray}
Then the mutual information reads
\begin{eqnarray}\label{e36}
\mathcal{I}\left( \hat{\rho} _{A}:\hat{\rho} _{B}\right)=2+\sum_{i=1}^{4}\lambda
_{i}\log_{2}\lambda_{i}.
\end{eqnarray}
Note that here we have used
$S(\hat{\rho}_{A}(t))=S(\hat{\rho}_{B}(t))=1$, since the two reduced
density matrixes $\hat{\rho}_{A}(t)$ and $\hat{\rho}_{B}(t)$ are
maximally mixed, that is $\hat{\rho}_{A}(t)=\hat{I}_{A}/2$ and
$\hat{\rho}_{B}(t)=\hat{I}_{B}/2$.

For calculation of the amount for the classical correlation
$\mathcal{C}(\hat{\rho})$ defined in Eq. (33), we propose the
complete set of orthogonal projectors
$\{\hat{P}_{k}=|\theta_{k}\rangle\langle\theta_{k}|,k=\parallel,\perp\}$
for a local measurement performed on the subsystem $B$, where the
two projectors are defined in terms of the following two orthogonal
states
\begin{eqnarray}
\label{e37}
|\theta_{\|}\rangle
&=&\cos\varphi|0\rangle+e^{i\phi}\sin\varphi|1\rangle,\nonumber\\
|\theta_{\perp}\rangle
&=&e^{-i\phi}\sin\varphi|0\rangle-\cos\varphi|1\rangle,
\end{eqnarray}
with $0\leq\varphi\leq\pi/2$ and $0\leq\phi\leq2\pi$. After the two
project measurements with $p_{\parallel}=p_{\perp}=1/2$, the reduced
density matrices of subsystem $A$ read as
\begin{eqnarray}
\label{e38} \hat{\rho} _{A}^{\parallel}&=&\frac{1}{4}\left(
\begin{array}{ll}
2(1+c_{3}\cos 2\varphi) & \epsilon \sin 2\varphi \\
\epsilon^{\ast} \sin 2\varphi & 2(1-c_{3}\cos 2\varphi)
\end{array}
\right), \nonumber\\
\hat{\rho} _{A}^{\perp}&=&\frac{1}{4}\left(
\begin{array}{ll}
2(1-c_{3}\cos 2\varphi) & -\epsilon \sin 2\varphi \\
-\epsilon^{\ast} \sin 2\varphi & 2(1+c_{3}\cos 2\varphi)
\end{array}
\right),
\end{eqnarray}
where we have introduced the parameter
\begin{equation}
\label{39} \epsilon=(c_{1}+c_{2})e^{-i\phi}+ c_0 e^{i\phi}.
\end{equation}

According to Eq. (37), it is straightforward to obtain the
eigenvalues of the reduced density matrix $\hat{\rho}_{A}^{k}$ as
follows
\begin{eqnarray}\label{e40}
\zeta^{1,2}_{\parallel}=\zeta^{1,2}_{\perp}=\frac{1}{2}(1\pm\Gamma),
\end{eqnarray}
where we have defined $\Gamma$ as
\begin{eqnarray}\label{e41}
\Gamma=\sqrt{c_{3}^{2}\cos^{2}(2\varphi)+\frac{|\epsilon|^{2}}{4}\sin^{2}(2\varphi)},
\end{eqnarray}
then we have
\begin{eqnarray}\label{e42}
S\left(\rho_{A}^{\parallel}\right)&=&S\left(\rho_{A}^{\perp}\right)
=f(\Gamma)\nonumber\\&=&-\frac{1+\Gamma}{2}\log_{2}\left(\frac{1+\Gamma}{2}\right)
-\frac{1-\Gamma}{2}\log_{2}\left(\frac{1-\Gamma}{2}\right),
\end{eqnarray}
which leads to the classical correlation
\begin{eqnarray}
\label{e43}
\mathcal{C}(\hat{\rho}(t))=1-\min_{\varphi,\phi}[f(\Gamma)].
\end{eqnarray}

Since the function $f(\Gamma)$ is a monotonically decreasing
function, in order to get the minimal value of $f(\Gamma)$  we
should choose proper parameters $\varphi$ and $\phi$ to ensure the
parameter $\Gamma$ defined in Eq. (40) is maximal. From Eqs. (38)
and (40) it is easy to get the following inequality
\begin{eqnarray}
\label{e44}
\Gamma&\leq&\sqrt{c_{3}^{2}\cos^{2}(2\varphi)+ c^2_4\sin^{2}(2\varphi)}\nonumber\\
&\leq&\left\{
\begin{array}{l}
\vert c_{3}\vert,\;\;\;\;\; $for$\;\;\;\vert c_{3}\vert
>c_4, \\
c_4,\;\;\;\;\;\;$for$\;\;\;\vert c_{3}\vert < c_4.
\end{array}\right.
\end{eqnarray}
where we have introduced the parameter
\begin{equation}
c_4=\frac{|c_{1}+c_{2}|+|c_0|}{2}.
\end{equation}

If we define $\gamma(t)$ as
\begin{eqnarray}
\label{e45} \gamma(t)=\max[\vert c_{3}\vert, c_4],
\end{eqnarray}
then the classical correlation can be expressed as
\begin{eqnarray}
\label{e46}
\mathcal{C}(\hat{\rho}(t))=\sum_{m=1}^{2}\frac{1+(-1)^{m}\gamma}{2}\log
_{2}[1+(-1)^{m}\gamma].
\end{eqnarray}
Therefore, the quantum discord can be written as
\begin{eqnarray}
\label{e47}
\mathcal{D}(\hat{\rho} (t))=2+\sum_{i=1}^{4}\lambda _{i}\log _{2}\lambda _{i}-\mathcal{C}(\hat{\rho}
(t)),
\end{eqnarray}
where the amount of the classical correlation $\mathcal{C}(\hat{\rho} (t))$ is given
by Eq. (46). In principle, we have obtained the dynamics of the
quantum discord according the above expression given in Eq. (47),
provided that we know the initial condition of the system. In what follows we
will study dynamic properties of the quantum entanglement and discord for some
initial states in detail.

From the density operator of the two DQD spin qubits at time $t$
(\ref{e28}) and Eq. (45) we can obtain the parameter
\begin{eqnarray}
\label{e48}
\gamma(t)=\left\{
\begin{array}{l}
\vert c_{3}\vert,\;\;\;\;\;\;$for$\;\;2|c_{3}|>|c_{1}-c_{2}|+|c_{1}+c_{2}|, \\
c_4,\;\;\;\;\;\;\;\;$for$\;\;2|c_{3}|<|c_{1}-c_{2}|e^{-2|\alpha|^{2}}+|c_{1}+c_{2}|.
\end{array}\right.
\end{eqnarray}
which indicates that the classical correlation expressed by Eq. (46)
largely depends on the initial state parameters of the two DQDs and
the TLR $\{c_1, c_2, c_3, \alpha\}$. Then making use of Eqs. (46-48)
we can obtain expressions of the quantum discord in different
regimes of the initial state parameters. When
$2|c_{3}|>|c_{1}-c_{2}|+|c_{1}+c_{2}|$, we have
\begin{eqnarray}
\label{49} \mathcal{D}(\hat{\rho}(t))&=&2+\sum_{i=1}^{4}\lambda
_{i}\log _{2}\lambda _{i}-\sum_{m=1}^{2}\frac{\gamma_{m}}{2}\log
_{2}\gamma_{m},
\end{eqnarray}
and when $2|c_{3}|<|c_{1}-c_{2}|e^{-2|\alpha|^{2}}+|c_{1}+c_{2}|$,
we have
\begin{eqnarray}
\label{50}
\mathcal{D}(\hat{\rho}(t))&=&2+\sum_{i=1}^{4}\lambda_{i}\log_{2}\lambda_{i}-\sum_{m=1}^{2}\frac{\gamma'_{m}}{2}\log_{2}\gamma'_{m},
\end{eqnarray}
where two parameters $\gamma_{m}$ and $\gamma'_{m}$ are defined by
\begin{eqnarray}
\label{51} \gamma_{m}&=&1+(-1)^{m}|c_{3}|,\;\;\;
\gamma'_{m}=1+(-1)^{m}c_4.
\end{eqnarray}

\begin{figure}[tbp]
\includegraphics[scale=0.62]{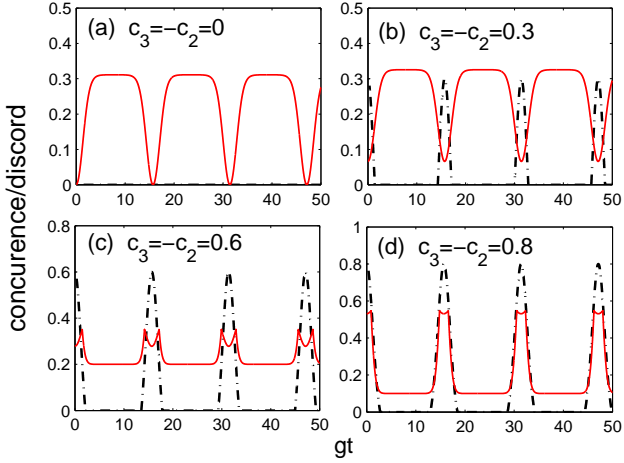}
\caption{(Color online)  Time evolution of concurrence (dot dashed
curves) and quantum discord (solid curves) for different values of
$c_{3}$ when other parameters are taken by $c_{1}=1, c_{2}=-c_{3},
\alpha=2, g=1,$ and $\delta=10$.} \label{fig2.eps}
\end{figure}

In order to see clearly dynamic characteristics of the quantum
entanglement and quantum discord, in the following we numerically
investigate the time evolution of the concurrence given by Eqs.
(\ref{e30}) and (\ref{e31}), the discord
$\mathcal{D}(\hat{\rho}(t))$ given by Eqs. (49) and (50).

Fig. 2 indicates the influence of initial states of the two DQD spin
qubits on dynamic evolution of the EC and DC when the DQD spin
qubits and the TLR are initial $X$-type quantum states and a
coherent state, respectively. We have plotted time evolution of
concurrence (dot dashed curves) and quantum discord (solid curves)
for some $X$-type initial quantum states  of the two spin qubits
when other parameters take fixed values. Fig. 2 reflects some
interesting dynamic properties of EC and DC.

\begin{figure}[tbp]
\includegraphics[scale=0.62]{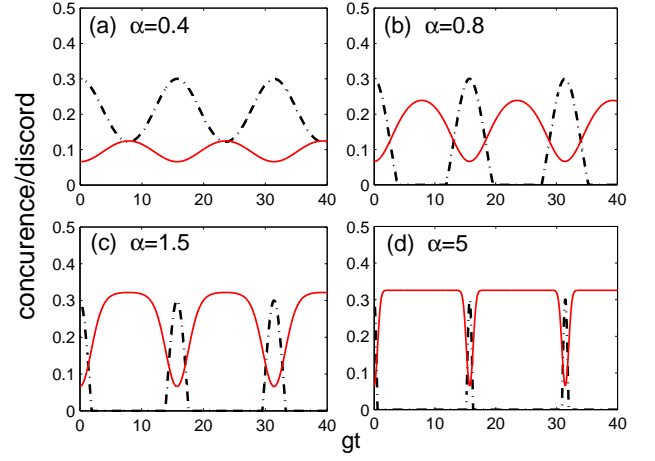}
\caption{(Color online)  Time evolution of concurrence (dot dashed
curve) and discord (solid curve)
 for different values of $\alpha$.
 Other parameters are $c_{1}=1, c_{2}=-c_{3}=-0.3, g=1,$ and $\delta=10$.}
\label{fig3.eps}
\end{figure}

(i) Fig. 2 demonstrates that in the EC death regions there exist
DC stationary states in which DC does not change in a finite time
area. From Fig. 2(a-d)  we can see that in the DC stationary
states the stable DC amplification or degradation can be generated
during the dynamical evolution. We can also see that these DC
stationary states can be controlled by initial-state parameters,
the coupling, and detuning between qubits and the TLR. Fig. 2(a-b)
clearly indicate that the DC may be amplified in the most time of
the death regime of the EC, and remains its maximal value.
However, Fig. 2(c-d) reflect the fact that the DC may be degraded
in the most time of the death regime of the EC, and remains its
minimal value. These imply that one can generate stable DC
amplification (degradation) in the EC death regions for certain
initial $X$-type states.

(ii) Fig. 2 indicates the appearance of the full synchronization
and anti-synchronization phenomena in the EC and DC time
evolution, and shows that the EC and DC synchronization
(anti-synchronization) depends on the initial-state parameters of
the two DQD spin qubits. Fig. 2(c) and Fig. 2(d) correspond to the
EC and DC synchronization evolution while Fig. 2(b) corresponds to
the EC and DC anti-synchronization evolution. From Fig. 2 we can
also see that both EC and DC time evolutions have the same period
which is independent of $c_2$ and $c_3$. Indeed, when $c_1=1$ and
$1>c_3=-c_2>0$ the evolution period of EC and DC can be
analytically obtained from Eqs. (28-30) with the simple expression
$T=\delta\pi/(2g)$. This implies that the larger period can
available by increasing (decreasing) the detuning $\delta$ (the
effective coupling constant $g$).

(iii) Fig. 2 indicates that both EC and DC may exhibit sudden
death phenomenon under certain conditions in their dynamic
evolution. Fig. 2(a) shows that the discord sudden death (DSD)
periodically occurs while EC always remains zero. From Fig. 2(b)
we can see that the entanglement sudden death (ESD) periodically
happens while DC almost remains its maximal value in the EC death
regions.

In Fig. 3 we show the effect of initial states of the TLR on EC
and DC dynamics. We have plotted time evolution of concurrence
(dot dashed curves) and quantum discord (solid curves) for some
initial quantum states of the TLR when other parameters take fixed
values. From Fig. 3 we can see that the time evolutions of EC and
DC are completely anti-synchronous with the same period, and the
ESD occurs for enough large values of the initial parameter
$\alpha$ in the EC dynamics. From Fig. 3(c) and Fig. 3(d) we can
also see that the increase of the initial-state parameter $\alpha$
of the TLR not only leads to the ESD  but also the appearance of
the DC stationary states in the EC death regimes. The lifetime of
these DC stationary states can be increased by increasing value of
the initial-state parameter $\alpha$.

In Fig. 4 we investigate the influence of the detuning parameter
on dynamic evolution of the EC and DC. We have plotted time
evolution of concurrence (dot dashed curves) and quantum discord
(solid curves) for different values of the detuning of the TLR for
a fixed initial state of the combined system and a fixed coupling.
From Fig. 4 we can see that the lifetime of the DC stationary
states and the EC death regimes increases with the increase of the
detuning parameter $\delta$. The larger the detuning, the longer
the lifetime of the DC stationary states and the EC death time
become. By using Eqs. (28-30), similar numerical analysis indicate
that the lifetime of the DC stationary states and the EC death
regimes can be manipulated by changing the coupling constant $g$.
Therefore, we can conclude that the lifetime of the DC stationary
states and the EC death regimes can be controlled by changing the
detuning parameter and the coupling constant between the DQD and
the TLR.

\begin{figure}[tbp]
\includegraphics[scale=0.62]{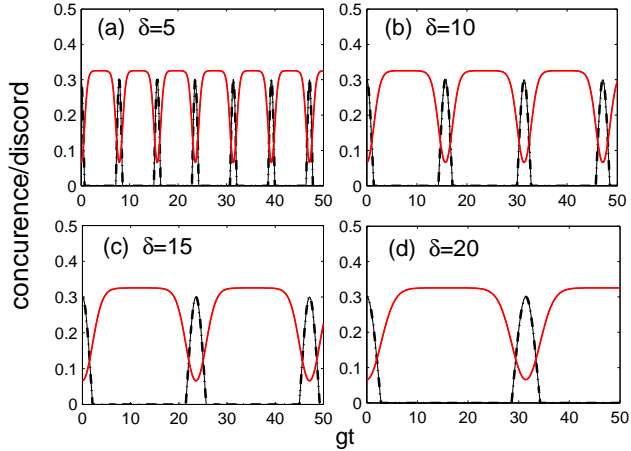}
\caption{(Color online)  Time evolution of concurrence (dot dashed
curves) and quantum discord (solid curves) for different values of
 $\delta$ when other parameters are taken by $c_{1}=1, c_{2}=-c_{3}=0.3,
\alpha=2$, and  $g=1$.} \label{fig4.eps}
\end{figure}

\section{\label{Sec:5} Concluding remarks}

In conclusion, we have proposed a magnetic coupling scheme between
two DQD spin qubits by using the TLR as a bus system. We have shown
that the inter-spin-qubit coupling is controllable and switchable.
The coupling controllability can be realized by changing external
magnetic field. It is worthwhile to mention that the present scheme
can be used to build a hybrid qubit system where DQD spin qubits are
integrated together with superconductiong qubits in the same TLR. We
have studied in some detail dynamical behaviors of inter-qubit
quantum correlations described by EC and DC when the two DQD spin
qubits and the TLR are initially prepared in some $X$-type quantum
states and a coherent state, respectively. We have studied the
relation between EC and DC. We have demonstrated that in the EC
death regions there exist DC stationary states.  In these DC
stationary states, the stable DC amplification or degradation can be
generated during the dynamical evolution. We have shown that the
lifetime of these DC stationary states depends on the initial-state
parameters, the coupling, and detuning between qubits and the TLR.
We have also found the full synchronization and anti-synchronization
phenomena in the time evolution of the EC and DC, and indicated that
the time synchronization (anti-synchronization) of the EC and  DC
dynamics depend on the initial-state parameters of the two DQD spin
qubits. We have indicated that both EC and DC may exhibit sudden
death phenomenon under certain conditions in their dynamic
evolution, and the DC stationary states always appear in the EC
death regions. The DC stationary states, synchronization and
anti-synchronization found in the present paper highlight new
characteristics in the EC and DC dynamics of the combined DQD-TLR
system, and have potential applications in quantum information
processing. It could be convinced that the magnetic coupling scheme
suggested in the present paper can be used to realize quantum
information processing since it takes advantages of controllable and
switchable coupling and qubit scalability.

\acknowledgments This work was supported by the National
Fundamental Research Program Grant No.  2007CB925204, the National
Natural Science Foundation under Grant Nos. 10775048 and 11075050,
the Program for Changjiang Scholars and Innovative Research Team
in University under Grant No. IRT0964, and the Research Fund of
Hunan Provincial Education Department Grant No. 08W012.


\end{document}